\documentclass[fleqn,usenatbib]{mnras}

\usepackage{newtxtext,newtxmath}

\usepackage[T1]{fontenc}
\usepackage{ae,aecompl}

\usepackage{graphicx}	
\usepackage{amsmath}	

\usepackage{subfig}

\usepackage{booktabs}

\usepackage{cleveref}
\crefname{section}{§}{§§}
\Crefname{section}{§}{§§}

\newcommand{\fstar}{$f_{*}$}

\newcommand{\Tto}{T$_{21}$}

\newcommand{\Lya}{Ly-${\alpha}$}

\newcommand*\mean[1]{\bar{#1}}

   \title[21-cm fluctuations from shot noise and scatter]{Shot noise and scatter in the star formation efficiency as a source of 21-cm fluctuations}

\author[I. Reis et al.]{
Itamar Reis$^{1}$\thanks{E-mail: itamarreis@mail.tau.ac.il},
Rennan Barkana$^{1}$,
and Anastasia Fialkov$^{2}$
\\
$^{1}$School of Physics and Astronomy, Tel-Aviv University, Tel-Aviv, 69978, Israel\\
$^{2}$Institute of Astronomy, University of Cambridge, Madingley Road, Cambridge CB3 0HA, UK\\
}

\date{Accepted XXX. Received YYY; in original form ZZZ}

\pubyear{2021}

\begin{document}
\label{firstpage}
\pagerange{\pageref{firstpage}--\pageref{lastpage}}
\maketitle

\begin{abstract}
   The 21-cm signal from cosmic dawn and the epoch of reionization (EoR) probes the characteristics of the high redshift galaxy population. Many of the astrophysical properties of galaxies at high redshifts are currently unconstrained due to the lack of observations. This creates a vast space of possible astrophysical scenarios where the 21-cm signal needs to be modeled in order to plan for, and eventually fit, future observations. This is done with fast numerical methods which make simplifying  approximations for the underlying physical processes. In this work we quantify the effect of Poisson fluctuations and scatter in the star formation efficiency; while Poisson fluctuations are included in some works and not in others, scatter in the star formation efficiency is usually neglected, and all galaxies of a given mass are assumed to have the same properties. We show that both features can have a significant effect on the 21-cm power spectrum, most importantly in scenarios where the signal is dominated by massive galaxies. Scatter in the star formation efficiency does not simply enhance the effect of Poisson fluctuations; for example we show that the power spectrum shape at cosmic dawn has a feature corresponding to the width of the galaxy brightness distribution. We also discuss some of the consequences for 21-cm imaging, and the signature of reduced correlation between the density and radiation fields.
  
\end{abstract}




\section{Introduction}

Measuring the redshifted 21-cm signal is one of the frontiers of observational astronomy, with the goal of providing a window to a currently unobserved era in the history of our Universe. Current observational efforts are focused on the time between cosmic dawn and the epoch of reionization (EoR), where
 the signal probes the properties of the first generation of stars and galaxies through their effects on the inter-galactic medium (IGM). 

  A first tentative measurement of the sky averaged (global) signal was recently reported \citep[EDGES, ][]{bowman18}, with many ongoing and future experiments aiming to test the EDGES result and, in addition, detect the 21-cm fluctuations across the sky. Such experiments  include the Shaped Antenna measurement of the background RAdio Spectrum \citep[SARAS,][]{patra13}, Probing Radio Intensity at high-Z \citep[PRIZM,][]{philip19},  the Large-Aperture Experiment to Detect the Dark Ages \citep[LEDA,][]{bernardi16},  the Mapper of the IGM Spin Temperature\footnote{http://www.physics.mcgill.ca/mist/} (MIST), and the Radio Experiment for the Analysis of Cosmic Hydrogen\footnote{https://www.kicc.cam.ac.uk/projects/reach} (REACH). Ongoing efforts to measure the fluctuations of the 21-cm signal include the Low Frequency Array \citep[LOFAR,][]{patil17}, the Murchison Wide-field Array \citep[MWA,][]{bowman13}, the GMRT-EoR experiment \citep[][]{paciga11}, the Owens Valley Radio Observatory Long Wavelength
Array \citep[OVRO-LWA,][]{eastwood19}, the Hydrogen Epoch of Reionization Array \citep[HERA,][]{deboer17}, LEDA  \citep{Hugh2021} and  the New Extension in Nan\c{c}ay Upgrading LOFAR \citep[NenuFAR,][]{zarka12}. Most of these experiments  have already produced first upper limits on the 21-cm power spectrum. In addition, the planned Square Kilometre Array \citep[SKA,][]{koopmans15} is expected to measure both the global signal and the power spectrum over a wide range of redshifts.

 The properties of the 21-cm signal are sensitive to the astrophysical parameters of the first stars and galaxies that formed in the Universe. Since the 21-cm signal will be the first to probe this epoch in cosmic history, these astrophysical parameters are virtually unconstrained. For this reason, fitting 21-cm measurements require fast simulation of the signal that can explore the vast space of possible astrophysical scenarios. Such exploration cannot be done with full cosmological simulations, and approximate methods that speed up the calculation are needed. An example for this was seen in the recent attempts at fitting the EDGES signal, with the use of various approximate techniques in \citet{mirocha19, fialkov19, ewall20, reis20c}.
 
 It is important to note that distinct methods include or neglect various astrophysical processes, and make non-identical astrophysical assumptions, which can lead to different constraints on the astrophysical parameters. For example, here are three differences between \citet{mirocha19} and \citet{reis20c} when fitting the EDGES signal: (i) \citet{mirocha19} assumed a star formation efficiency that increases with time, while \citet{reis20c} assumed it to be constant. (ii) \citet{mirocha19} assumed a fixed SED for X-rays emitted by the first galaxies, while in \citet{reis20c} the X-ray SED was left as a free parameter. (iii) Heating of the IGM by \Lya{} heating was neglected in \citet{mirocha19} but included in \citet{reis20c}. These differences and others resulted in  \citet{mirocha19} and \citet{reis20c} obtaining different constraints on the X-ray production efficiency when fitting the same data.
 
 We expect the issue of different model assumptions to be significant in future attempts to fit the fluctuations of the 21-cm signal across the sky. In this work we focus on two such issues: 
 
 (i) Poisson fluctuations in the number of galaxies. Some methods, including our previous work \citep{visbal12, fialkov14a, cohen16}, calculated the expectation value for the number of halos in each pixel of the simulation given analytical models \citep{press74, sheth99, barkana04}. Such methods neglect the shot-noise contribution, i.e., the contribution of Poisson fluctuations, to the number of halos, which can have a signature in the resulting 21-cm signal. Other works calculated a halo field composed of individual halos \citep{mesinger07, baek09, santos10, zahn11}, and as a result did include the Poisson fluctuations. We note that some of these methods focused only on the EoR, and/or were restricted to only include massive halos. In \citet{reis20a} we added Poisson fluctuations to our simulation and discussed the observational possibilities of individual halos during cosmic dawn, focusing on the coupling transition. In this work we explore the effect of Poisson fluctuations on the 21-cm signal at various stages of its evolution. 
 
 (ii) Scatter in the star formation efficiency. To the best of our knowledge, all of the existing semi-numerical 21-cm calculations assume that all galaxies of a given mass have the same astrophysical parameters. On the other hand, full simulations as well as observations imply that galaxies, even of the same mass, show scatter in their properties. For example, \citet{xu16} reported a roughly log-normal distribution of the star formation efficiency in \texttt{Renaissance Simulations} galaxies. The width of the distribution depends on the mass of the galaxy, with $\sigma(f_{*}) \lesssim 2$. While  \citet{xu16} present results only for $z \leq 15$, they provide a motivation for investigating the effect of \fstar{} scatter at higher redshifts. In this work we explore the effect of such scatter on 21-cm fluctuations, adopting a scatter in the star formation efficiency as an example.

\section{Simulating the 21-cm signal}
\label{sec:methods}
In a schematic picture of the evolution of the 21-cm signal, the radiation emitted by the first sources generated the following three cosmic 21-cm events \citep{madau97, barkana18book, Mesingerbook}: (i) The coupling transition, in which stellar \Lya{} photons coupled the spin temperature (describing the occupation levels of the hydrogen ground state hyperfine levels) and the kinetic temperature of the IGM, through the Wouthuysen-Field effect \citep[WF effect, ][]{wouthuysen52, field58}. This breaks the coupling between the spin temperature and the temperature of the background radiation (which we assume is the cosmic microwave background, CMB, though there can also be a contribution from additional high redshift radiation sources); this enables the detection of a 21-cm signal, which can only be seen relative to the CMB. (ii) Heating of the IGM, most likely due to X-ray emission, for example from X-ray binaries. Before heating takes place the kinetic temperature is lower than the CMB temperature, resulting in a 21-cm signal seen in absorption. If  the kinetic temperature of the gas rises above the CMB, the 21-cm signal is seen in emission. (iii) Reionization, in which UV radiation ionizes the IGM and gradually eliminates the 21-cm signal (which is only emitted from neutral hydrogen).

The brightness temperature of the 21-cm signal, relative to the CMB, can be approximated by the following equation \citep{barkana18book}, for an optical depth $\tau \ll 1$:
\begin{multline}
\label{eqn:tb}
T_{\rm 21}  =  26.8 \left(\frac{\Omega_{\rm b} h}{0.0327}\right)
\left(\frac{\Omega_{\rm m}}{0.307}\right)^{-1/2}
\left(\frac{1 + z}{10}\right)^{1/2} \\
(1 + \delta) x_{\rm HI} \frac{x_{\rm tot}}{1 + x_{\rm tot}} \left( 1 - \frac{T_{\rm rad}}{T_{\rm K}}\right) [{\rm mK}],
\end{multline}
where $\delta$ is the baryon overdensity, $x_{\rm HI}$ is the neutral fraction, $x_{\rm tot}$ is the coupling coefficient between the spin temperature and the CMB temperature, and $T_{\rm K}$ is the kinetic temperature of the gas. Also $T_{\rm rad} = T_{\rm CMB} + T_{\rm Radio}$, where $T_{\rm Radio}$ is a possible contribution from high redshift radiation sources, namely galaxies \citep{mirocha19,ewall20, reis20c} or exotic physics \citep{fialkov19}.  Note that in the simulation itself we use the full equation for $T_{\rm 21}$, which is non-linear in $\tau$.

The 21-cm signal thus reflects the imprint on the IGM of the radiation fields produced by high redshift galaxies. The sky averaged (global) signal is sensitive to the strength of the various fields, and probes mean properties such as the overall star formation rate (SFR) and the efficiencies for producing the various radiation fields. In addition to the global signal, the fluctuations of the 21-cm signal across the sky contain rich information about the clustering properties of high redshift galaxies,  the spectrum of the radiation sources, and the density fluctuations of the IGM. In this work we explore the 21-cm signatures of some of the properties of the high redshift galaxy population.

We use our own  semi-numerical code to calculate the 21-cm signal \citep[e.g.,][]{visbal12, fialkov14, cohen17, fialkov19}. The simulation computes the population of galaxies at each redshift, as we describe in detail below. Given the galaxy field, the various radiation fields emitted by these galaxies are obtained. The effect of these radiation fields on the IGM is then computed. Namely, the simulation finds the following IGM properties needed for the evaluation of the 21-cm signal: the kinetic temperature of the gas, the coupling coefficient, the spin temperature and the neutral fraction. In this work we use a 128$^3$ grid  with a resolution of 3 comoving Mpc. The outputs of the simulation are cubes of the 21-cm brightness temperature at each redshift. From these we calculate the global signal and the spherically-averaged power spectrum. The code was originally inspired by  \texttt{21cmFAST} \citep{mesinger11}, but has a completely independent implementation.

\subsection{The high redshift galaxy population}
\label{sec:pressschechter}

The input of the simulation is a realization of the initial density and velocity fields \citep[calculated using power spectra generated with the publicly available code  \texttt{CAMB},][]{camb}. The density and velocity fields are evolved using linear perturbation theory.
 Given the linear density field, the mean number of dark matter halos $n(M,z)$ at a given mass and redshift is calculated with the hybrid approach of \citet{barkana04} which combines the previous models of  \citet{press74} and \citet{sheth99}. This hybrid model was shown to give a good match to simulations in regions of various sizes and mean densities. 

Given a single dark matter halo of mass $M$, the mass of baryons that are contained in the halo is obtained with the help of fits to hydrodynamical simulations \citep{tseliakhovich11, fialkov12}. Such a halo will contain stars if $M > M_{\rm cool}$, where $M_{\rm cool}$ is a minimum mass for gas cooling and star formation, and is a parameter of the simulation. We parametrize this minimum mass for star formation by the circular velocity, defined as the velocity of a circular orbit at the virial radius of the halo. For a halo of mass $M$:
\begin{multline}
    V_c = 16.9 \left(\frac{M}{10^8 M_{\odot}}\right)^{1/3}\left(\frac{1+z}{10}\right)^{1/2} \\ \left(\frac{\Omega_m h^2}{0.141}\right)^{1/6} \left(\frac{\Delta_c}{18 \pi^2}\right)^{1/6}  [{\rm km}\;{\rm s}^{-1}]\ ,
\end{multline}
where $\Delta_c$ is the ratio between the collapsed  density and the critical density at the time of collapse, and equals $18 \pi^2$ for spherical collapse. In addition to this cutoff, our simulation  includes the suppression of  star formation due to the relative velocity between dark matter and gas \citep{tseliakhovich10,fialkov12}, Lyman-Werner feedback \citep{haiman97,fialkov2013} and photoheating feedback \citep{rees86, sobacchi13,cohen16}. In the simulation, galaxies form stars with a star formation efficiency parametrized by \fstar.

\textbf{Poisson fluctuations:}
\label{sec:poisson_procedure}
Calculating the number of halos as described above works well when there is a large number of halos in a simulation pixel. For small numbers of halos it fails to capture the shot noise coming from the fact that halos are integer quantities. In such a case a realization of the actual number of halos can be computed by drawing the number of halos from a Poisson distribution, where the mean of the distribution is the prediction of the model for the number of halos.

To take into account the Poisson fluctuations in the number of halos we treat the number of halos \textit{created in each time step of the simulation} as a Poisson variable. In each time step we calculate the mean number of halos created in the time step, by subtracting the mean number of halos in the beginning from the number of halos at the end  of the time step. The final number of halos created in the time step is then  drawn  from a Poisson distribution. This is done separately for each mass bin, as well as each pixel. We create all the halos in the simulation using this Poisson procedure. 

We note  that our procedure does not take into account correlations between the number of halos in different mass bins.  In reality the number of halos in different mass bins is correlated, for example, by the process of mergers, in which two halos of a low mass 'disappear', and are replaced by a single halo of a higher mass. Our procedure is accurate in the early stage of
galaxy formation, when most pixels have one star-forming halo or none, and so the correlations have no effect. Our procedure is also accurate later on, once the Poisson fluctuations become negligible and the halo mass distribution reverts to the mean. Thus, our results in this paper are most reliable in the 
early and late stages; there may be a lower accuracy in the intermediate regime, which we plan to check in future work.

\textbf{Scatter in the astrophysical parameters:}
The parameters of the simulation include physical properties of halos such as the star formation efficiency. Usually in a semi-numerical simulation, all halos (of a given mass) have the same physical properties. In reality different halos can have different physical properties. This can be the result of various effects such as different environments or merger histories.

In this work we take advantage of the individual halos created through the Poisson procedure, and consider the effect of a variance in their physical properties. Since all the types of radiation from galaxies that affect the 21-cm signal are proportional to the star formation efficiency \fstar{}, we focus on this key parameter; also, we focus on the early stage of cosmic dawn when the 21-cm signal is driven by the \Lya{} intensity (before the X-ray or ionizing backgrounds have been built up substantially), and thus \fstar{} is the observationally-important efficiency parameter. For each new halo we draw $\log_{10}{f_{*}}$  from a normal distribution (i.e., the \fstar{} distribution is assumed to be log-normal). The variance is left as a free parameter of the simulation.  Since \fstar{} is an efficiency, we want to always keep $f_{*} \leq 1$. To do this, given a variance and a desired mean value $\mean{f}_{*}$, we numerically find a correction factor so that after multiplying our log-normal distribution by that factor and truncating at unity, we indeed get the desired mean value.  

\subsection{The radiation fields}
Given a population of galaxies, we calculate the various radiation fields and their effect on the IGM.  The \Lya{} radiation field is responsible for the WF coupling, and \Lya{} photons also contribute to the heating of the IGM. We assume that the intensity of the \Lya{} radiation of a given galaxy is proportional to its star formation rate (SFR). The X-ray luminosity is also assumed to be proportional to the SFR \citep[e.g.,][]{Mineo:2012, fragos13, fialkov14a, Pacucci:2014}, but with an additional efficiency factor $f_{X}$, where $f_X = 1$ corresponds to low-metallicity galaxies observed at low redshifts. The X-ray spectral energy distribution (SED) is assumed to have a power law shape with a slope $\alpha$ and low-frequency cutoff $\nu_{\rm min}$.  Reionization is implemented using the excursion set formalism \citep{furlanetto04}, with  $\zeta$ representing the ionizing efficiency of sources, and R$_{\rm mfp}$ the maximum distance ionizing photons can travel (due to Lyman limit systems).

The free parameters of the simulation are thus: \fstar,  $V_c$,  $f_{X}$, $\alpha$, $\nu_{\rm min}$,  $\zeta$,  R$_{\rm mfp}$. As described above, in this work we add an additional parameter corresponding to the variance in \fstar. Except for \fstar, all other simulation parameters are kept constant for all galaxies in the simulation, though in principle it is of course possible that these other astrophysical parameters will show significant galaxy to galaxy variance as well.

\section{Results}
\label{sec:case_study}

\subsection{The coupling transition}

\begin{figure}
    \centering
     \includegraphics[width=0.99\columnwidth]{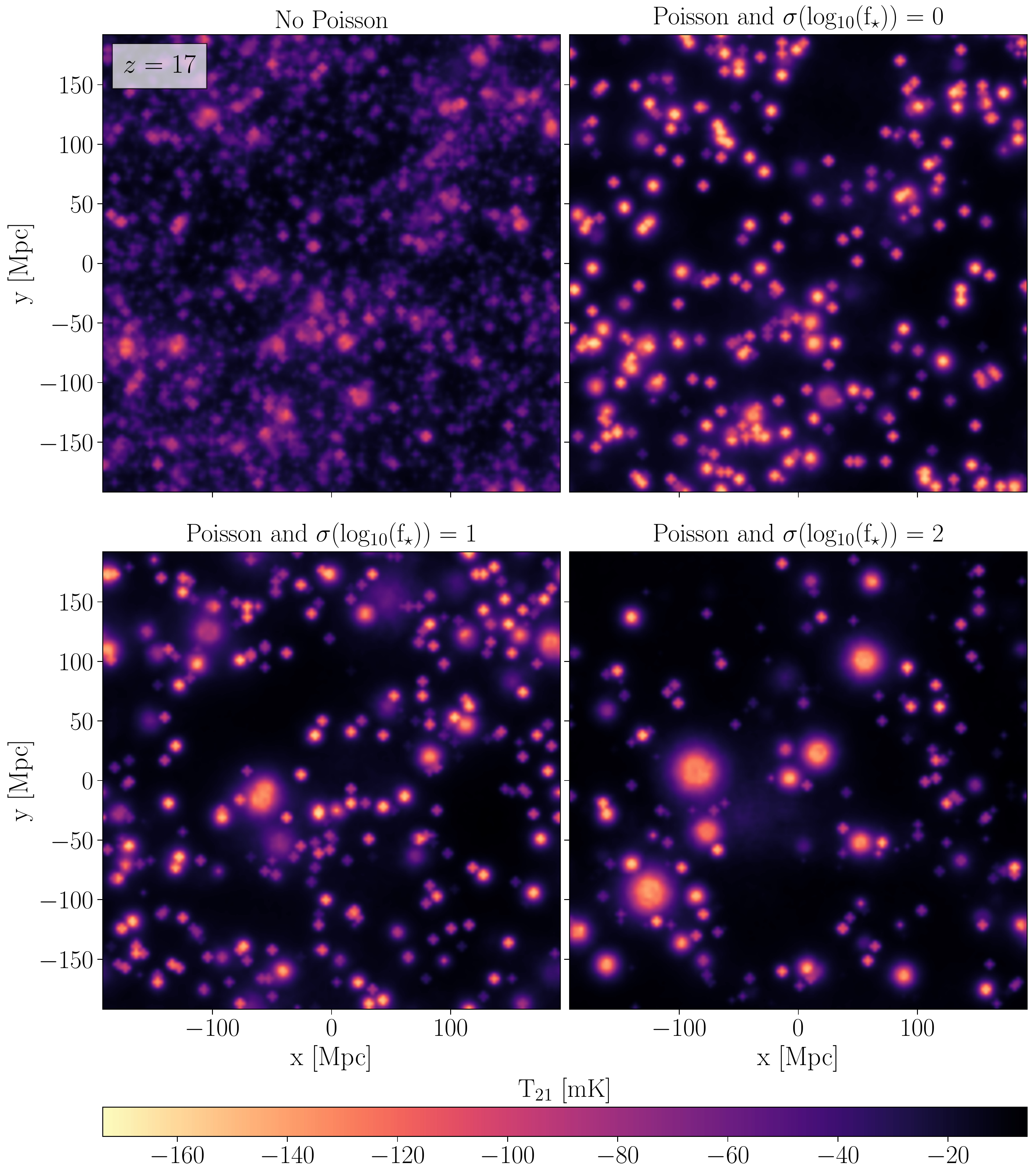}   
    \caption{Projected images of the 21-cm signal during the coupling transition for models with increasing size of scatter in the star formation efficiency. Since early galaxies in this model were rare, we find it useful to show a kind of projected image, defined as showing the minimum value of the signal in the direction perpendicular to the image (obtained from a simulation box that is 384 comoving Mpc on a side; each image is made of square pixels of side 3 Mpc). We note that a similar projection can be easily done with actual future data. We show an astrophysical model with $V_c$ = 50 km s$^{-1}$, and mean \fstar{} = 0.01. }
    \label{fig:example_coupling_slices}
\end{figure}

In order to understand the effect of scatter, we begin with a visual inspection of 21-cm images from our simulations. Fig.~\ref{fig:example_coupling_slices} shows projected images of the 21-cm signal during the coupling transition ($z = 17$) for an astrophysical model with $V_c$ = 50 km s$^{-1}$, and \fstar{} = 0.01. We show the "No Poisson" case (i.e., without Poisson fluctuations and also without \fstar{} scatter), the case with Poisson fluctuations but without \fstar{} scatter, and two cases with Poisson fluctuations and increasing \fstar{} scatter.

The projected images show the minimum values of the signal (deepest absorption) along the axis perpendicular to the image. In the case without Poisson fluctuations, \Lya{} radiation is produced by every pixel in the simulation, according to the predicted mean number of galaxies  of the modified extended Press-Schechter model. Importantly, in a given pixel, this predicted number can be significantly smaller than one, and the pixel will still produce non-zero radiation. The strongest observed 21-cm absorption occurs at density peaks, where the mean number of galaxies predicted by our  model is the largest (note: in the specific model shown here, with relatively large galaxies, the streaming velocity plays a negligible role). This results in a strong \Lya{} radiation field in these regions, which produce the absorption signal through the WF effect. In the case with Poisson fluctuations the 21-cm signal is much more localized, i.e., confined to only a fraction of the overall volume. Each bubble seen in the top right panel of Fig.~\ref{fig:example_coupling_slices} corresponds to an individual galaxy, where the \Lya{} radiation produced by the galaxy creates a "coupled bubble" around it. Bubbles of different sizes correspond to galaxies of different brightness. Since in the top panel of Fig.~\ref{fig:example_coupling_slices} every galaxy has the same \fstar, the different brightness is only due to the mass of the galaxy. Note that the projected images show all the galaxies in our 3 dimensional simulation box.

In the bottom panels of Fig. \ref{fig:example_coupling_slices} we introduce  \fstar{} scatter. Now, different galaxies of the same mass can have different star formation efficiencies, and thus different values of \Lya{} brightness. This creates a wider distribution of bubble sizes, as can be seen in the images. The larger the size of the scatter in \fstar, the wider the distribution of bubble sizes. The large bubbles in the tail of the distribution are very important for 21-cm imaging as they may be directly detected by the SKA \citep{reis20a}. Thus, 21-cm imaging is highly sensitive to \fstar{} scatter; while in the top right panel of Fig. \ref{fig:example_coupling_slices} no bubble is large enough to be individually detected, a model with the same astrophysical parameters except high \fstar{} scatter (bottom right panel of Fig. \ref{fig:example_coupling_slices}) does produce a number of large, likely detectable, individual bubbles.

\begin{figure}
    \centering
     \includegraphics[width=0.99\columnwidth]{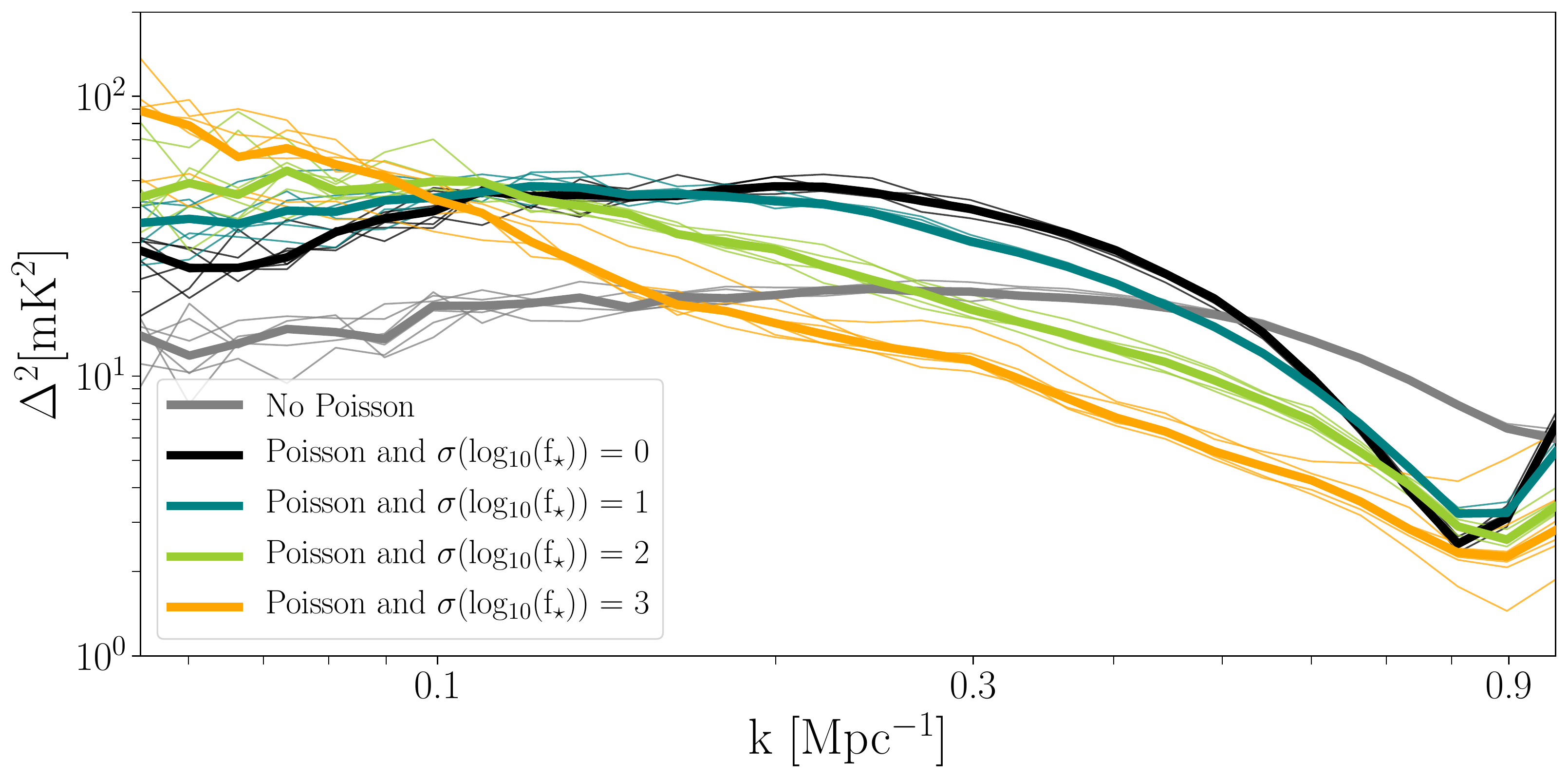}

    \caption{The 21-cm power spectrum shape during the \Lya{} coupling transition, for models with increasing size of scatter in the star formation efficiency. Thin lines show individual simulation runs and thick lines show the median power spectrum of the individual runs. We assume the same astrophysical model as in Fig.~\ref{fig:example_coupling_slices}.}   
    \label{fig:coupling_peak_ps}
\end{figure}

Fig. \ref{fig:coupling_peak_ps} shows the power spectrum shape at the coupling peak (defined as the redshift where the power spectrum at $k = 0.1$ Mpc$^{-1}$ peaks due to fluctuations in the coupling field),  for the same astrophysical parameters as in Fig. \ref{fig:example_coupling_slices}. Comparing the power spectrum shape for the cases with and without Poisson fluctuations (gray and black lines, respectively), we see that Poisson fluctuations enhance 21-cm fluctuations on large scales (small $k$), and reduce the power spectrum on small scales (large $k$). This is in contrast to the naive expectation that the contribution of Poisson fluctuations will behave like $k^3$ (shot-noise power spectrum, multiplied by $k^3$ in the standard convention used in our plots), that is, more power on large $k$. The naive expectation is only valid on scales significantly larger than the size of the coupled bubbles. The break seen in the power spectrum (referring to the black line, the case with Poisson fluctuation) thus corresponds to the typical size of the coupled bubbles:
\begin{equation}
    k_{\rm break} = \frac{2 \pi}{R_{\rm bubble}}\ .
\end{equation}
On larger scales, where the 21-cm fluctuations are dominated by the spatial distribution of galaxies, we see an enhancement of the 21-cm fluctuations due to Poisson fluctuations. On smaller scales, below the size of the coupled bubbles (which are larger due to the imposed minimum unit of one galaxy, and act as a smoothing scale for the 21-cm image), we see a suppression of the 21-cm fluctuations.

Introducing \fstar{} scatter affects this picture by making the distribution of bubble sizes wider. This can be seen in Fig. \ref{fig:coupling_peak_ps} by inspecting the change in the power spectrum shape when the size of the scatter in \fstar{} is increased. For the case with Poisson fluctuations but no \fstar{} scatter, the coupled bubbles all have similar sizes, corresponding to a relatively well defined $k_{\rm break}$. For cases with increased variance in \fstar{} and a wider distribution of bubble sizes, $k_{\rm break}$ is no longer well defined, and the transition between large scale enhancement and small scale suppression occurs over a wide range of $k$'s. This shows that \fstar{} scatter does not result in a simple enhancement of the effect of Poisson fluctuations, and can potentially produce a unique signature in the power spectrum shape. Indeed, in the limit of very large scatter, the fluctuations monotonically decreases with $k$, which is sensible since if we consider two points separated by some distance, the smaller the distance, the higher the probability that if one of them is in a \Lya{} bubble then the other point is in the same bubble, and the 21-cm intensity is nearly the same at both points.

\subsection{Heating transition}
\begin{figure}
    \centering
     \includegraphics[width=0.99\columnwidth]{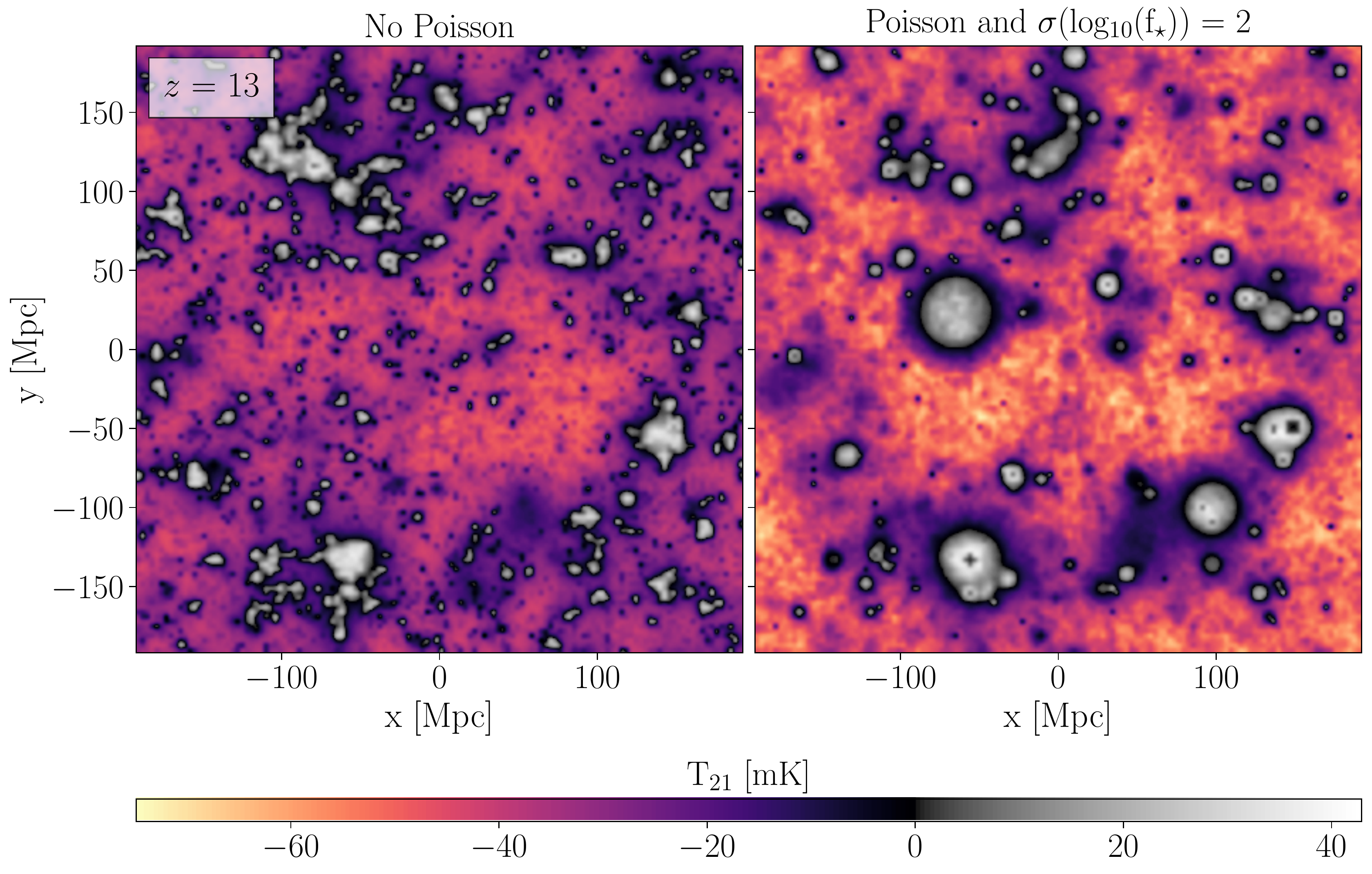} \\[6pt]
     \includegraphics[width=0.99\columnwidth]{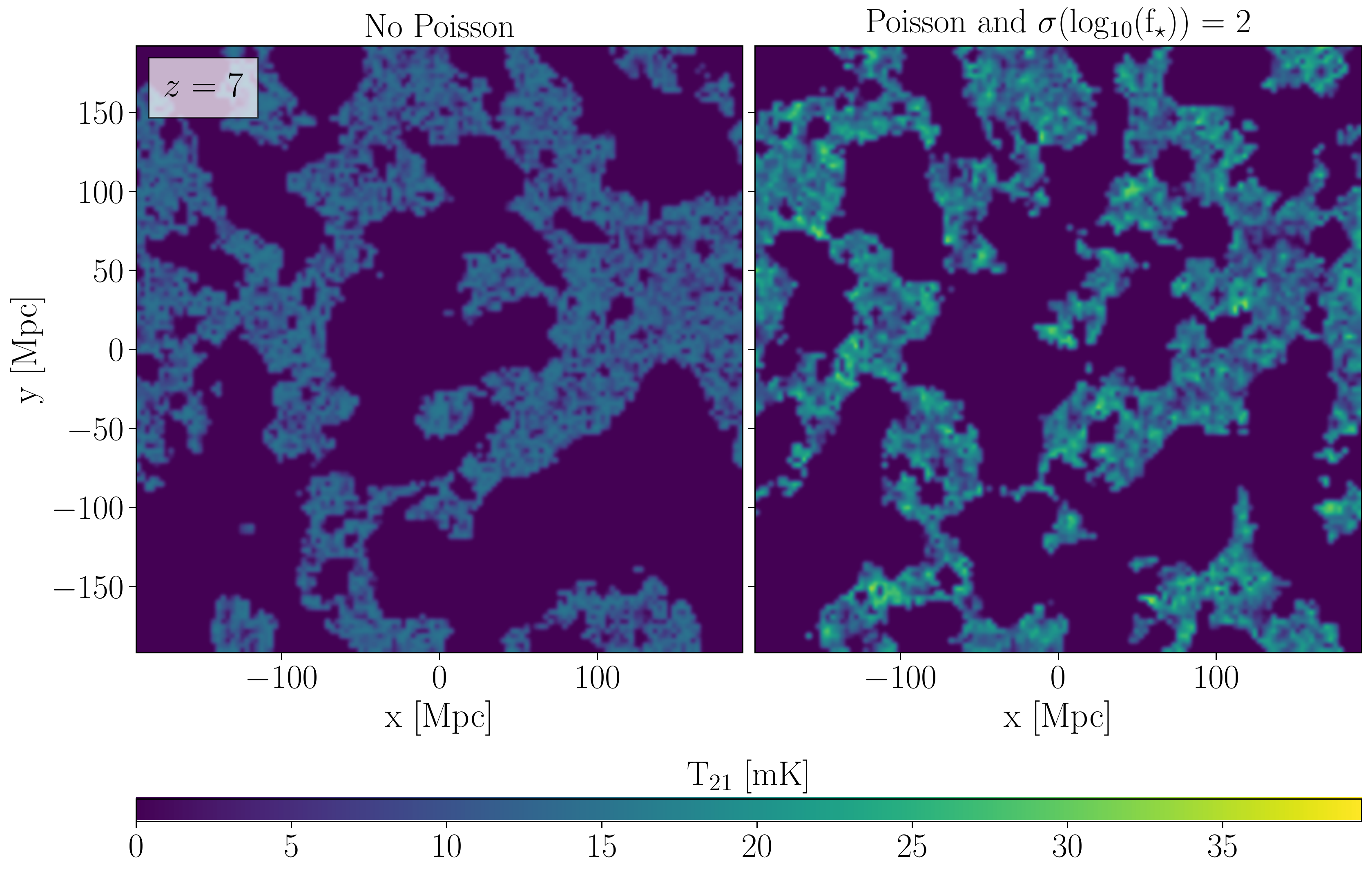}
    \caption{Images of the 21-cm signal during the heating ({\bf top panel}) and reionization ({\bf bottom panel}) transitions. As in the previous figures, we assume an astrophysical model with $V_c$ = 50 km s$^{-1}$ and mean \fstar{} = 0.01. In this example we show a case with efficient X-ray heating; $f_X = 10$ and a relatively soft SED (minimum X-ray energy $\nu_{\rm min}$ = 0.1 keV).  }
    \label{fig:example_heating_and_ion_slices}
\end{figure}

During the heating transition the effect of Poisson fluctuations and \fstar{} scatter depends on the properties of the X-rays that produce the heating. Soft X-rays travel short distances, and produce heating near their sources. Thus, in models dominated by soft X-rays, the heating signature of individual galaxies can be seen in the 21-cm signal. This scenario is shown in the upper panels of Fig. \ref{fig:example_heating_and_ion_slices}. Roughly spherical "heated bubbles" are produced around individual galaxies.

\begin{figure}
    \centering
     \includegraphics[width=0.99\columnwidth]{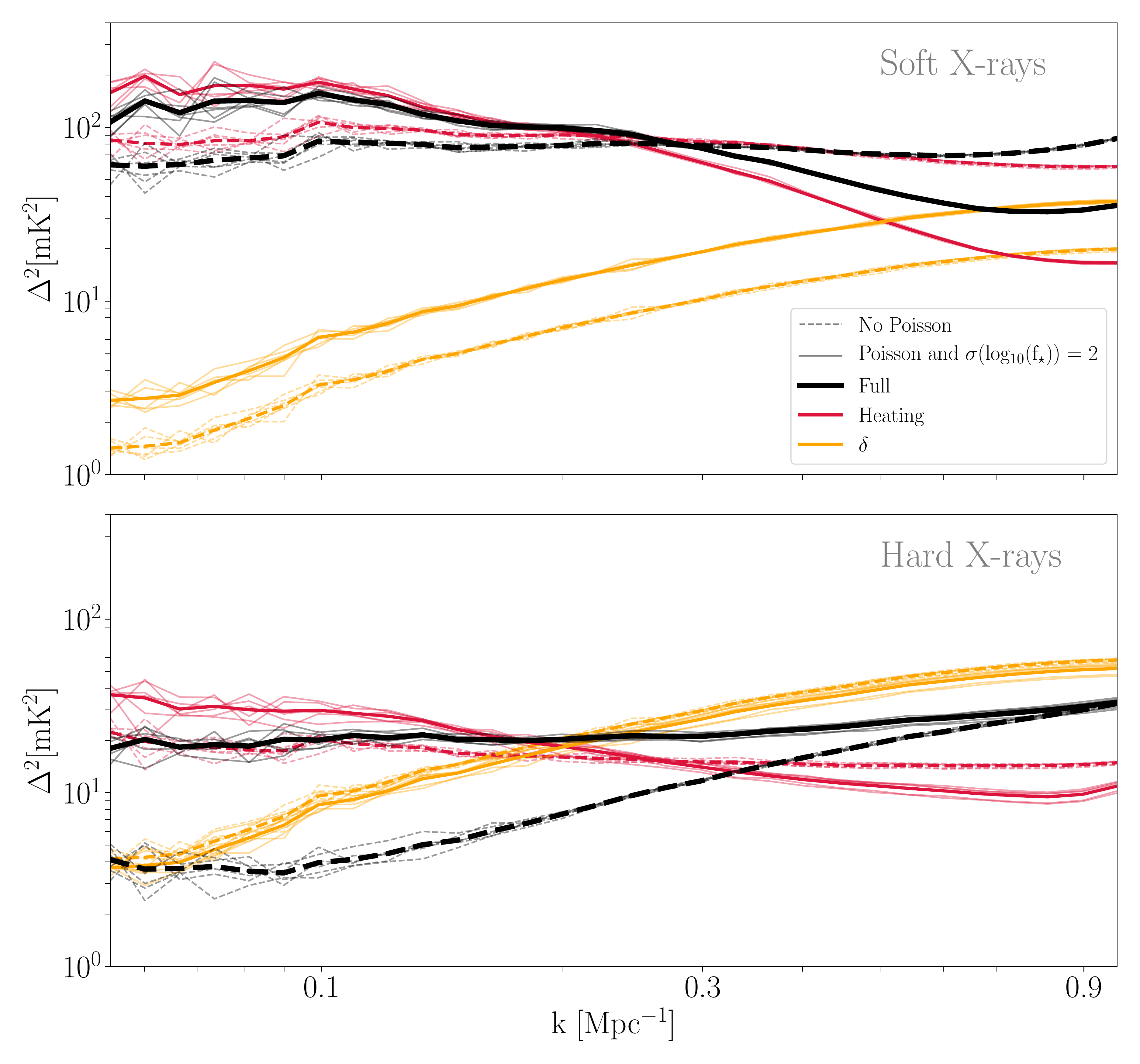}   
    \caption{The 21-cm power spectrum at the heating peak (maximum heating-induced 21-cm fluctuations) for an astrophysical model with soft X-rays ({\bf top panel}) or hard X-rays ({\bf bottom panel}). In each panel we show the power spectrum with Poisson fluctuations and large variance in the star formation efficiency (solid lines), and compare to a model without Poisson fluctuations or \fstar scatter (dashed lines). We also show the separate contributions to the power spectrum (in a linear approximation) from heating fluctuations and from density fluctuations. Thin lines show individual runs of the simulation, while thick lines show the median power spectrum of these runs. The rest of the astrophysical parameters are as in Fig.~\ref{fig:example_heating_and_ion_slices}; in particular, the top panel shows the same model as in the top panel in Fig.~\ref{fig:example_heating_and_ion_slices}. }
    \label{fig:heating_peak_ps}
\end{figure}

Fig. \ref{fig:heating_peak_ps} shows the power spectrum shape at the heating peak for two astrophysical models that differ in their X-ray SED; one with a relatively soft SED ($\nu_{\rm min}$ = 0.1 keV) and one with a hard SED ($\nu_{\rm min}$ = 1 keV). In both cases we compare the power shape without Poisson fluctuations to a model with both Poisson fluctuations and large \fstar{} scatter. For the case of soft X-rays, heating fluctuations dominate 21-cm fluctuations at the heating peak, while density fluctuations are relatively small (except at the high-$k$ end of the range of scales that we consider). In this case the change in shape of the power spectrum at the heating peak is similar to the change in shape observed at the coupling peak (Fig.~\ref{fig:coupling_peak_ps}, with a large scale enhancement and small scale suppression. Here, this is due to heated bubbles around individual galaxies producing a similar effect on the power spectrum as the coupled bubbles discussed above. Again, the scale of the break in the power spectrum correspond to the typical size of the bubbles. Note that the heated bubbles can indeed be seen in the top panels of Fig.~\ref{fig:example_heating_and_ion_slices}.

For hard X-rays, which travel much larger distances before absorption by the IGM, the heating transition is much more spatially uniform, and the power spectrum is significantly lower \citep{fialkov14, fialkov14a}. In this case individual galaxies do not produce such a strong effect on the nearby IGM, and the effect of Poisson fluctuations and \fstar{} scatter is smaller, though its scale-dependence is similar in terms of the effect on the power spectrum contribution from heating fluctuations. Since in this case heating is relatively uniform, 21-cm fluctuations during the heating transition are not dominated solely by heating fluctuations, and density fluctuations also play a substantial role. Thus, in the example shown here the power spectrum is strongly enhanced at small $k$ and barely affected at high $k$, resulting in an overall flattened slope. Importantly, since density and heating fluctuations are of similar strength, the  fact that their contributions to 21-cm fluctuations are anti-correlated has an important effect on the final signal. Poisson fluctuations and \fstar{} scatter decrease this anti-correlation, as the locations of galaxies (which are the X-ray sources, and heat the gas around them) become less well-correlated with density peaks. For example, at $k > 0.3$ Mpc$^{-1}$, while the separate linear contributions are both reduced, the final power spectrum is actually enhanced, due to the strong reduction in the anti-correlation.  This is an example of a general effect of Poisson fluctuations and \fstar{} scatter, namely the suppression of correlation or anti-correlation between density and the various radiation fields. The effect of this on the final 21-cm power spectrum is significant if and when the 21-cm fluctuations receive substantial contributions from both density fluctuations and another source of fluctuations (that is, coupling, heating or reionization).

\subsection{Reionization}

An example of the effect of Poisson fluctuations on  reionization is shown in the bottom panels of Fig.~\ref{fig:example_heating_and_ion_slices}. Two effect can be seen; some difference in the location and sizes of the ionized regions, and the value of the 21-cm field in the neutral regions. The first difference comes from the Poisson fluctuations in the spatial distribution of galaxies and the fluctuations in their SFR. The second difference is due to the effect of Poisson fluctuations on the neutral fraction outside the ionized bubbles. Without Poisson fluctuations we have significant partial ionization in pixels outside the ionized bubbles, while with Poisson fluctuations the neutral fraction is close to unity in these regions (X-ray induced ionization slightly decreases the neutral fraction). This is the cause of the stronger 21-cm signal outside the ionized bubbles which is clearly seen in the bottom panels of Fig.~\ref{fig:example_heating_and_ion_slices}. In simulation runs without Poisson fluctuations, pixels that contain a low, below unity mean number of halos, partially self ionize and possibly also affect their close surroundings. Adding Poisson fluctuations makes pixels without at least one galaxy empty, and then the inside-out topology of reionization emanates from the star-forming halos outwards, as expected.

\subsection{Dependence on $V_c$ and \fstar}
Poisson fluctuations and  \fstar{} scatter are most important in models and epochs  where the radiation fields are dominated by a small number of rare galaxies. For a given model, Poisson fluctuations and \fstar{} variance will be most important at high redshift, and their significance will decrease with time as more galaxies form. It is possible, for example, that Poisson fluctuations and \fstar{} variance will have a significant effect on the 21-cm signal during cosmic dawn, and become negligible during the EoR.

In addition, Poisson fluctuations and \fstar{} scatter are very important in some models, but can be neglected in others. The most important astrophysical parameters to determine this are the circular velocity threshold $V_c$ (corresponding to the minimum mass for star formation), and the (mean) star formation efficiency \fstar{} (as well as other radiation production efficiencies). Poisson fluctuations are more important for models with higher $V_c$, in which there are fewer star-forming galaxies at a given redshift. Poisson fluctuations are also more important for models with higher radiation production efficiencies, in which, for a given $V_c$, the cosmic 21-cm milestones occur at earlier redshifts, when there were fewer star-forming galaxies. 

\begin{figure*}
    \centering
     \includegraphics[width=0.99\textwidth]{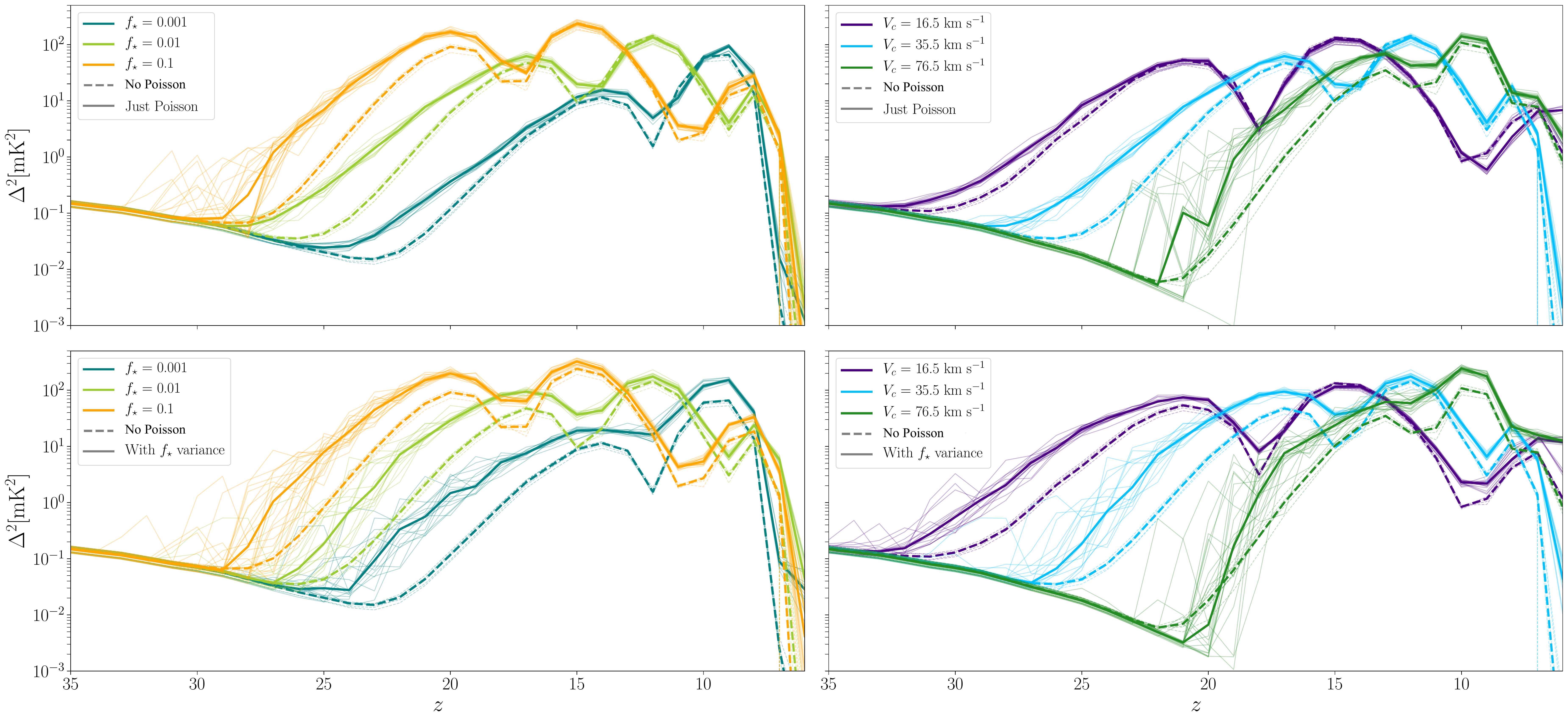}

    \caption{The power spectrum at $k = 0.1$ Mpc$^{-1}$ as a functions of redshift. {\bf Top panels:} The effect due to Poisson fluctuations alone. {\bf Bottom panels:} The effect due to Poisson fluctuations and variance in \fstar{}, for the case of $\sigma(\log_{10}(f_{*})) = 2$.  {\bf Left panels:} Fixed $V_c = 35.5$ km s$^{-1}$, and various values of \fstar. {\bf Right panels:} Fixed \fstar = 0.01 and various values of $V_c$. For all models we assume $f_X = 1$ and a soft X-ray SED ($\nu_{\rm min}$ = 0.1 keV), and also compare to the case with no Poisson fluctuations or \fstar{} variance.}
    \label{fig:power_spectrum_vs_z}
\end{figure*}

The top panels of Fig. \ref{fig:power_spectrum_vs_z} show some examples of the evolution of the power spectrum at $k = 0.1$ Mpc$^{-1}$, with and without Poisson fluctuations. In all cases shown we see that the size of the effect induced by Poisson fluctuations and \fstar{} scatter  is  strongest during cosmic dawn and can be seen even for the model with the lowest-mass galaxies considered ($V_c = 16.5$ km s$^{-1}$). For this model, the effect of Poisson fluctuations becomes negligible by the redshift of the "coupling peak" (the highest-redshift peak of the power spectrum, which is due to fluctuations in the coupling field). For higher values of $V_c$, the effect of Poisson fluctuations can be seen down to lower redshifts, but in all cases the effect is strongest during cosmic dawn.

In the bottom panels of Fig. \ref{fig:power_spectrum_vs_z} we add the effect of \fstar{} variance. In all cases, the combined effect of Poisson fluctuations and \fstar{} variance is significantly stronger (at some redshifts) than for "just Poisson". For example, for the \fstar{} = 0.01, $V_c$ = 35.5 km s$^{-1}$ model, at the coupling peak ($z \sim 17$), the power spectrum is enhanced by a factor of 1.5 by Poisson fluctuations alone, and by a factor of 2.6 by the combined effect of Poisson fluctuations and \fstar{} variance.
We note that for the model with a high mean value of \fstar{} = 0.1 (orange lines), the truncation of the \fstar{} values at \fstar{} = 1 effectively leads to a reduced \fstar{} variance.

It is also interesting to discuss the run-to-run variance, which is related to the cosmic variance. For each case shown in Fig.~\ref{fig:power_spectrum_vs_z} we ran multiple independent runs of the simulation. For each simulation we used a different realization of the initial density field, independent draws of halos from the Poisson distribution (in the cases with Poisson fluctuations), and independent draws of \fstar{} (in the cases with \fstar{} variance). We ran more simulations for cases with high run-to-run variance, with twenty four simulation runs for cases with the highest run-to-run variance and six for cases with the lowest run-to-run variance.

In the top panels of Fig. \ref{fig:power_spectrum_vs_z} it can be seen that the run-to-run variance is larger for the case with Poisson fluctuations compared to the case without, especially at high redshifts. For example, for the \fstar{} = 0.01, $V_c$ = 35.5 km s$^{-1}$ model, the run-to-run variance for the case with Poisson fluctuations is substantially larger than in the case without Poisson fluctuations, at the beginning of cosmic dawn ($z \sim 25$). This mean that at this epoch, the variance due to Poisson fluctuations strongly dominates over the variance due to the density field. But already at the coupling peak ($z \sim 17$) the run-to-run variance becomes similar for both cases, which means that the density field becomes an important source of variance. 

The bottom panels of Fig. \ref{fig:power_spectrum_vs_z} show that introducing variance in \fstar{} significantly increases the run-to-run variance, suggesting that this source of variance can dominate over Poisson fluctuations. For example, for the \fstar{} = 0.01, $V_c$ = 35.5 km s$^{-1}$ model, at $z\sim 25 - 20$, the run-to-run variance for the case with \fstar{} variance is much larger than in the case with only Poisson fluctuations. We note that to directly relate the run-to-run variance to the observed cosmic variance, one would need to take into account the size of the simulation box compared to the field of view of the survey of interest. For example, an SKA field of view at $z=20$ with a bandwidth of 10~MHz corresponds to roughly 18 of our simulation volumes. 

\section{Summary}
\label{sec:summary}

We have explored the effect of Poisson fluctuations in the number of halos and of scatter in the star formation efficiency on the fluctuations of the 21-cm signal:
\begin{enumerate}
    \item{{\bf Individual galaxy effects on 21-cm images and the 21-cm power spectrum.} Taking Poisson fluctuations into account allows us to obtain the signatures of individual galaxies in the 21-cm signal. Such signatures can in many cases be directly seen in 21-cm images, as coupled, heated, or ionized bubbles around individual galaxies. Individual galaxies can also produce an effect on the 21-cm power spectrum. For example, during the coupling transition, a break in the power spectrum appears, corresponding to the typical size of coupled bubbles around individual galaxies. 
    
    Similarly, the \fstar{} scatter affects both 21-cm images and the power spectrum. By creating a tail of bright galaxies that produce a strong signature, high \fstar{} scatter enhances the possibility of observing the signature of individual galaxies in the 21-cm signal. In terms of the power spectrum, we have shown that \fstar{} scatter can produce enhancement on large scales on top of that of Poisson fluctuations, and even in cases where the contribution from Poisson fluctuations alone is small (Fig.~\ref{fig:power_spectrum_vs_z}). As far as the power spectrum shape at cosmic dawn (Fig. \ref{fig:coupling_peak_ps}), \fstar{} scatter does not produce a simple enhancement of Poisson fluctuations, as the power spectrum shape shows a signature of the {\it width} of the galaxy brightness distribution.   }
    \item{{\bf Suppression of the correlation between the density and radiation fields.} This is a more subtle effect that becomes important whenever the 21-cm fluctuations receive comparable contributions from the density field and from one of the radiation fields. In such a case, the total 21-cm fluctuations are strongly affected by the spatial correlation between the density field and the relevant radiation field. This correlation is weakened by Poisson fluctuations  and by \fstar{} scatter due to the introduction of a random component to the spatial distribution of the radiation field.}
    \item{{\bf Enhancement of cosmic variance.} Both Poisson fluctuations and \fstar{} scatter increase the run-to-run variance in the power spectrum, with a significant increase due to \fstar{} scatter, even in cases when the Poisson increase is small.}
    
\end{enumerate}

We note again that we used scatter in \fstar{} as an example for the general effect of scatter in the astrophysical parameters on 21cm fluctuations. Future work can explore the effect of scatter in other parameters.

\section*{Acknowledgments}

This project was made possible for I.R. and R.B. through the support of the Israel Science Foundation (grant No. 2359/20) and the ISF-NSFC joint research program (grant No. 2580/17). AF was supported by the Royal Society University Research Fellowship. 

This research made use of:
 {\fontfamily{cmtt}\selectfont  SciPy} \citep[including {\fontfamily{cmtt}\selectfont  pandas} and {\fontfamily{cmtt}\selectfont NumPy, }][]{2020SciPy, numpy},  {\fontfamily{cmtt}\selectfont IPython} \citep[][]{perez07}, {\fontfamily{cmtt}\selectfont matplotlib} \citep[][]{hunter07}, {\fontfamily{cmtt}\selectfont Astropy} \citep[][]{astropy-collaboration13}, {\fontfamily{cmtt}\selectfont Numba} \citep{lam15},  the SIMBAD database \citep[][]{wenger00}, and the NASA Astrophysics Data System Bibliographic Services.

\section*{Data availability}
No new data were generated or analysed in support of this research.

\bibliographystyle{mnras}
\bibliography{fluctuations}

\bsp	
\label{lastpage}
\end{document}